# The effect of collisions in ionogram inversion

*Carlo Scotto[a,b], Alessandro Settimi[a]*

[a] Istituto Nazionale di Geofisica e Vulcanologia, Via di Vigna Murata 605, 00143, Rome, Italy

[b] Doctoral School in Polar Sciences, University of Siena, Via del Laterino 8, 53100, Siena, Italy

*Corresponding author:*

email: alessandro.settimi@ingv.it; phone: +390651860719; fax: +390651860397

**Abstract**

The results of this paper demonstrate how the effect of collisions on the group refraction index is small, when the ordinary ray is considered. If, however, in order to improve the performance of a system for automatic interpretation of ionograms, the information contained in ordinary and extraordinary traces is combined, the effect of collisions between the electrons and neutral molecules should be taken into account for the extraordinary ray.

## 1. Introduction

The problem of obtaining an electron density profile from an ionogram is fundamental for the physics of the ionosphere. This importance was perceived very early, and even Budden's historical book (1961) addresses the issue extensively, discussing a solution based on polynomial methods. Titheridge (1985) again proposed polynomial methods and, on the strength of newly available computing power, the Polan (Polynomial Analysis) program was introduced for the manual inversion of ionogram traces. Similarly, the ARTIST system (Automatic Real Time Ionogram Scaler with True height), created by Huang and Reinish (1996), also applies a polynomial technique in which an electron density profile $N_e(h)$ is expressed for each ionospheric layer using shifted Chebyshev polynomials. More recently a technique was proposed in which an ionogram trace is computed using a probable electron density profile, minimizing the root mean square error between the restored and the recorded trace (Scotto, 2009). This approach is very similar to the function approximation methods used in many branches of applied mathematics. These methods generally resolve problems of approximation by selecting a particular function belonging to a well-defined class that closely approximates a target function (hence the name of the method) in a task-specific way. This computation is performed by the Adaptive Ionospheric Profiler, which is an integral part of Autoscala, the software developed at the National Institute of Geophysics and Vulcanology Roma,– Italy, for the automatic scaling of ionograms (Scotto and Pezzopane, 2002; Pezzopane and Scotto, 2004).

In all these methods, it is seen that with appropriate assumptions for the valley of the E region, if present, it is possible to derive an electron density profile from either ordinary or extraordinary traces.

In order to improve the accuracy of the estimate of $N_e(h)$, recently Scotto et. al. (2012) have introduced a numerical method to improve the accuracy of the calculation by which the ionogram trace is reconstructed. With this method, the time calculation is immediately optimized. The refractive index which was used in that work took into account the presence of the Earth's magnetic field and neglected the effect of electron-neutral molecule collisions.

In this paper, we show how this assumption is in fact justified. Precisely, it is shown with numerical examples, that between the ionograms reconstructed starting from the same $N_e(h)$, using collisional and collisionless refraction indices, with magnetic field, no appreciable differences are observable.

## 2. The method of the target function

As stated above, in the method of the target function it is necessary to iteratively perform the calculation of the ionogram trace from an electron density profile. For this purpose, it is necessary to calculate the following integral:

$$h'(f) = \int_0^{h_R} \mu_g [f, f_p(h)] \cdot dh, \qquad (1)$$

where $f$ is the frequency of the radio wave, $f_p$ is the plasma frequency at height $h$, and $h_R$ is the height at which reflection takes place. In general, $\mu_g$ is the real part of the group refractive index $n_g$, which is a complicated function of $f$, $f_p$, $f_b$ and dip angle. For an ordinary ray in the integral (1) it is involved the real part $\mu_{g[ORD]}$ of group refractive index $n_{g[ORD]}$ and the height $h_R$ is the value at which $f_p(h_R)=f$. Furthermore, as it is well note, $\mu_{g[ORD]}$ can be calculated from the Appleton-Hartree equation. Neglecting the collisions and considering the presence of the Earth's magnetic field **B**, after some mathematical manipulations (Shinn and Whale, 1951), in the frequency range of interest here ($f_p<f$), it is:

$$\mu_{g[ORD]} = -\frac{1}{2f\mu_{f[ORD]}} \left\{ \frac{1}{f^2 D} \left[ 2f_p^2 f - \left(f^2 \mu_{f[ORD]}^2 - f^2\right) \frac{Y_L^2 f\left(1-X^2\right)}{\sqrt{\frac{1}{4}Y_T^4 + Y_L^4\left(1-X\right)^2}} \right] + 2f \right\}, \quad (2)$$

where the respective refractive index of phase $\mu_{f[ORD]}$, neglecting collisions and considering the presence of **B**, is:

$$\mu_{f[ORD]}^2 = 1 - \frac{X(1-X)}{D}, \quad (3)$$

with:

$$D = (1-X) - \frac{Y^2 \sin^2\theta}{2} + \sqrt{\frac{Y^4 \sin^4\theta}{4} + Y^2(1-X)\cos\theta} \quad (4)$$

and:

$X=\omega_p^2/\omega^2$ ($\omega_p=Ne^2/m\varepsilon_0$ is the plasma frequency, $\omega$ is the angular frequency of the radio wave, $N$ is the electron density, $m$ is the electron mass, $\varepsilon_0$ is the constant permittivity of vacuum);

$Y=\omega_B/\omega$ ($\omega_B=Be/m$ is the girofrequency, and $B$ the magnetic field amplitude of the Earth);

$\theta$ is the angle between the wave vector and the magnetic field of the Earth.

Thus ignoring collisions, the Adaptive Ionospheric Profiler (AIP) (Scotto, 2009) estimates a profile of electron density based on the ordinary trace, using a refractive index that takes into account the presence of the magnetic field but without collisions. Of course, the same electron density profile could be obtained by a similar procedure, using the appropriate refractive index $\mu_{g[EXT]}$, from the extraordinary ray. Even $\mu_{g[EXT]}$ can be expressed by (2), but instead of (4) it is necessary to use:

$$D = (1-X) - \frac{Y^2 \sin^2\theta}{2} - \sqrt{\frac{Y^4 \sin^4\theta}{4} + Y^2(1-X)\cos\theta}. \qquad (5)$$

## 3. The effect of collisions on ionogram traces

The group refractive index with collisions can be calculated using the following equation:

$$n_g = n_f + \frac{\partial n_f}{\partial \omega}\omega, \qquad (6)$$

which is valid for any dispersive medium. Since the real part $\mu_g$ of $n_g$ needs to be calculated, the expression can be better specified as:

$$\mu_g = \mu_f + \frac{\partial \mu_f}{\partial \omega}\omega. \qquad (7)$$

In (7), instead of the generic $\mu_f$, considering the phase refraction index for an ordinary ray with collisions $\mu_{fc[ORD]}$ (or an extraordinary $\mu_{fc[EXT]}$), a corresponding group refraction index $\mu_{gc[ORD]}$ (or $\mu_{gc[EXT]}$) is obtained. The phase refraction indices $\mu_{fc[ORD]}$ and $\mu_{fc[EXT]}$ can be derived from the well-known Appleton-Hartree equation. Once $\mu_{fc[ORD]}$ and $\mu_{fc[EXT]}$ are known, from (7) it is easy to obtain $\mu_{gc[ORD]}$ (or $\mu_{gc[EXT]}$) by calculating $\partial \mu_{fc[ORD]}/\partial \omega$ (or $\partial \mu_{fc[EXT]}/\partial \omega$) numerically.

Subsequently, an ionospheric electron density profile was approximately modelled by a parabolic layer, which is widely used as it simply fits the shape of the peak of the maximum electron density $N_mF2$ observed in F2 region (Davies, 1990). This parabolic layer starts from the altitude of 110 km, and has a maximum $N_0 = 2 \; 10^{12}$ m$^{-3}$ at an altitude of 380 km.

A direction of radio wave propagation was hypothesized at an angle $\theta = 30°$ to the Earth's magnetic field, and the intensity of the latter was assumed sufficient to cause a girofrequency $f_B$=1.2 MHz, compatible with the ionograms recorded at the ionospheric station in Rome. For the collision frequency it was also assumed a rough model based on the figure published by Davies (1990) in which $\upsilon$ varies with height as $\upsilon = 10^{-0.0125 \cdot h[km]+5.5}$ [s$^{-1}$]. On the basis of these assumptions, virtual height versus frequency graphs were calculated, simulating ionograms, assuming two different values of maximum electron density in the ionosphere for both the ordinary and extraordinary rays. Figures 1 (a)-(b) shows the results obtained for the ordinary ray, and Figures 2 (a)-(b) the corresponding results for the extraordinary ray.

These figures show that collisions have a slightly appreciable effect on the heights of reflection of the extraordinary ray, so making it interesting to introduce them in the case we would estimate the electron density profile from the extraordinary trace.

**4. The estimation of the electron density profile from an ordinary trace using refractive indexes with and without collisions**

As mentioned in section 2, it is possible to estimate the vertical profile of electron density with the method of the target function. In Autoscala program, this method is used from the recorded ordinary trace, by making use of the refractive index $\mu_{g[ORD]}$, in which collisions are neglected.

On the other hand, the same operation can be carried out by making use of a refractive index in which, instead, it is taken into account that there are collisions between the electrons and neutral molecules. In Figure 3(a), is shown, by way of example, one ionogram where the vertical profile of electronic density was estimated using $\mu_{g[ORD]}$, while in Figure 3(b), however, the inversion was performed using $\mu_{gc[ORD]}$. It is seen that the parameters provided by Autoscala are the same, for both the indices of refraction that can be used.

# 5. Reconstruction of an extraordinary trace from an ordinary trace using refractive indexes with and without collisions

Section 2 above describes how AIP can provide electron density for altitude $N_e(h)$ by adjusting the parameters of an analytical model, producing a profile from which an artificial ionogram is obtained, matching the observed ordinary trace. Using a corresponding refraction index $\mu_{g[EXT]}$ for the extraordinary ray, the same profile could be used within (1) to obtain an appropriate reconstruction of the extraordinary trace.

In Figure 4(a) is reported to the magnification of an ionogram where the extraordinary trace restoring was performed using $\mu_{g[EXT]}$, while in Figure 4(b) the same restoring was carried out using $\mu_{gc[EXT]}$. In this case, a more realistic model is used for the collision frequency of electron-neutral molecules.

Specifically it is assumed (Rishbeth, 1969) that:

$$\upsilon(h) = 5.4 \cdot 10^{-16} \cdot n(h) \cdot T(h)^{1/2} \text{ [m.k.s.]},$$

where for $n(h)$ and $T(h)$ simplified models with parabolic trends were used to appropriate transition heights with linear development thereafter. It is observed a slight but significant difference between the traces that are obtained, which is in agreement with the results of numerical simulations illustrated in section 3 and in Figures 2 (a)-(b).

# 6. Conclusions

The results of the numerical simulations shown in Figures 1(a)-(b) and 2(a)-(b) demonstrate how the effect of collisions on the group refraction index is small, when we consider the ordinary ray. In the ionograms inversion, therefore, since one usually makes use of the information contained in the recorded ordinary trace, it is convenient to use a collisionless refraction index.

If, however, in order to improve the performance of a system for automatic interpretation of ionograms, we wanted to combine the information contained in ordinary and extraordinary traces, the effect of collisions between the electrons and neutral molecules should be taken into account, for the extraordinary ray. Note, however, that for $\mu_{gc[EXT]}$, it is difficult to obtain an analytical expression, so its calculation is carried out through the numerical evaluation of the expression (7). This assessment is possible, but results in an increasing of computation time.

**References**

Budden, K. G., Radio waves in the ionosphere, Cambridge University Press, Cambridge, United Kingdom, 1961.

Davies, K., Ionospheric Radio, Peter Peregrinus Ltd. (ed.), London, UK, 1990.

Huang, X., Reinisch, B. W., Vertical Electron density profiles from the digisonde network, Adv. Space Res. 18 (6), 121-(6)129, 1996.

Pezzopane, M., Scotto, C., Software for the automatic scaling of critical frequency foF2 and MUF(3000)F2 from ionograms applied at the Ionospheric Observatory of Gibilmanna, Ann. Geophysics, 47, 1783-1790, 2004.

Rishbeth, H., Garriott, O. K., Introduction to ionospheric physics, Academic Press, New York, USA, 1969.

Scotto, C., Electron Density profile calculation technique for Autoscala ionogram analysis, Adv. Space Res. 44, 756-766, 2009.

Scotto, C., Pezzopane, M., A software for automatic scaling of foF2 and MUF(3000)F2 from ionograms, Proceedings of URSI 2002, Maastricht, 17-24 august, 2002 (on CD).

Scotto, C., Pezzopane, M., and Zolesi, B., Estimating the vertical electron density profile from an ionogram: On the passage from true to virtual heights via the target function method, Radio Scie.Vol. 47, RS1007, 6 PP., 2012, doi:10.1029/2011RS004833.

Shinn, D. H., Whale, H. A., Group velocities and group heights from magnetoionic theory, J. Atmos. Terr. Phys. 2, 85-105, 1952.

Titheridge, J. E., Ionogram analysis with the generalised program POLAN, World Data Center A for Solar-Terrestrial Physics, Boulder, CO (USA), Technical Report PB-89-152482/XAB UAG-93, 204 pp., 1985.

**Figure captions**

**Figure 1 (a)-(b).** Simulated ionograms for a simplified ionosphere in which $N_e(h)$ is modelled by a parabolic layer. The ordinary trace is simulated assuming an index of refraction that does not take collisions into account (in red), and an index including collisions (in blue). A frequency of collisions varying with altitude was assumed at $\upsilon= 10^{-0.0125 \cdot h[km]+5.5}$ [s$^{-1}$]. It is seen that the effect of collisions does not appear to significantly affect the shape of an ordinary trace.

**Figure 2 (a)-(b).** Simulated ionograms for a simplified ionosphere in which it is assumed that $N_e(h)$ varies in a parabolic shape. The extraordinary trace is simulated assuming an index of refraction that does not take collisions into account (in red), and an index including collisions (in blue). A frequency of collisions varying with altitude was assumed at $\upsilon= 10^{-0.0125 \cdot h[km]+5.5}$ [s$^{-1}$]. It is seen that the effect of collisions slightly affects the shape of the extraordinary trace, all over the HF frequencies range.

**Figure 3 (a)-(b).** Ionograms in which $N_e(h)$ was estimated from the ordinary trace of the ionogram making use of the group refractive index $\mu_{g[ORD]}$ where collisions are not considered (a) and a refractive index $\mu_{gc[ORD]}$ group where, instead, they are taken into account of (b). The frequency of collisions was assumed varying with altitude as $\upsilon= 10^{-0.0125 \cdot h[km]+5.5}$ [s$^{-1}$]. No appreciable differences are observable between the two sets of parameters obtained by Autoscala.

**Figure 4 (a)-(b).** An ionogram in which the reconstruction of the extraordinary trace was carried out starting from the same $N_e(h)$ estimated by the recorded ordinary trace by making use of (a) $\mu_{g[EXT]}$ and (b) $\mu_{gc[EXT]}$.

**Figure 1**

(a)

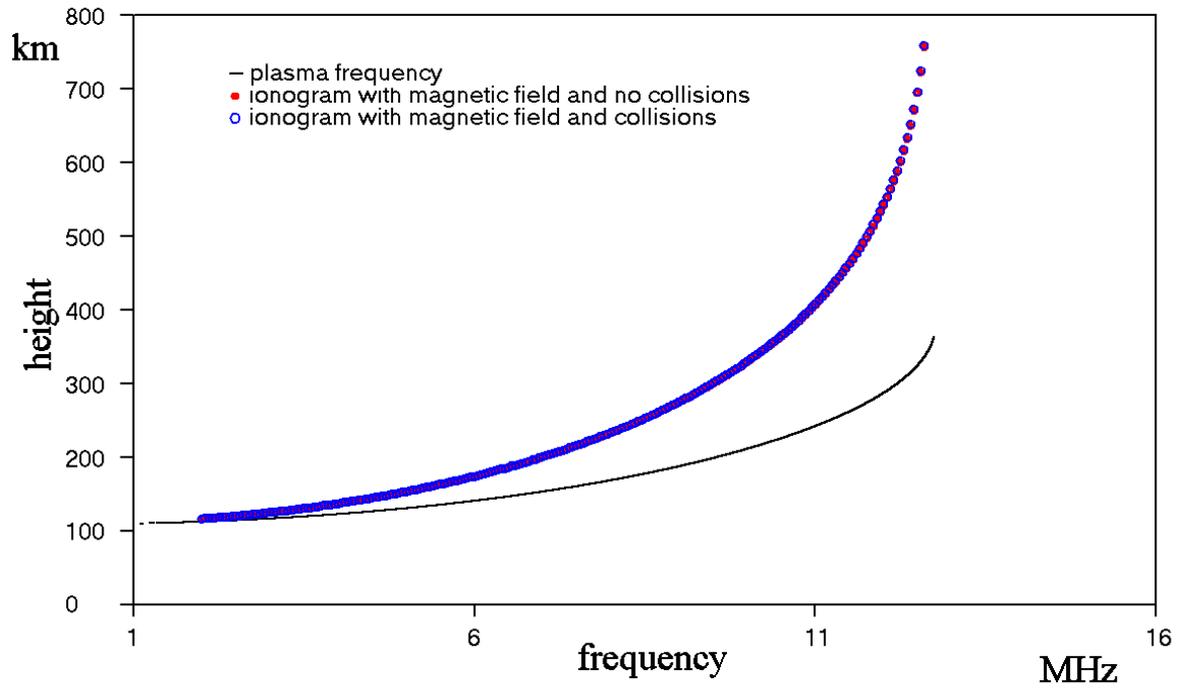

(b)

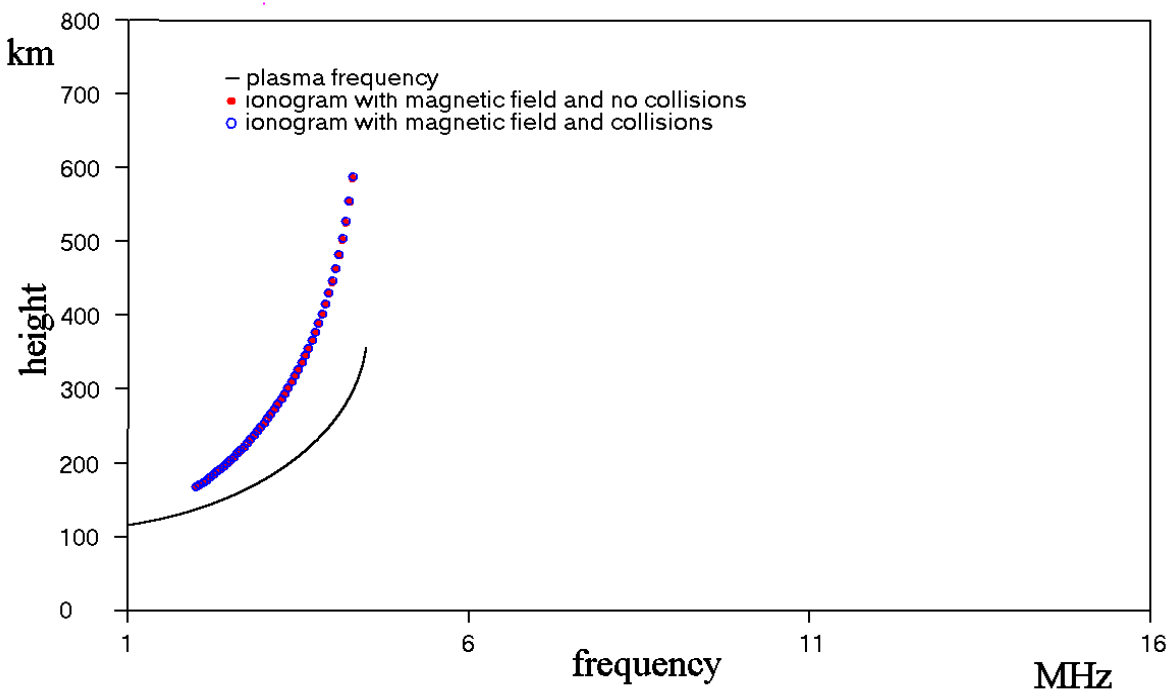

**Figure 2**

(a)

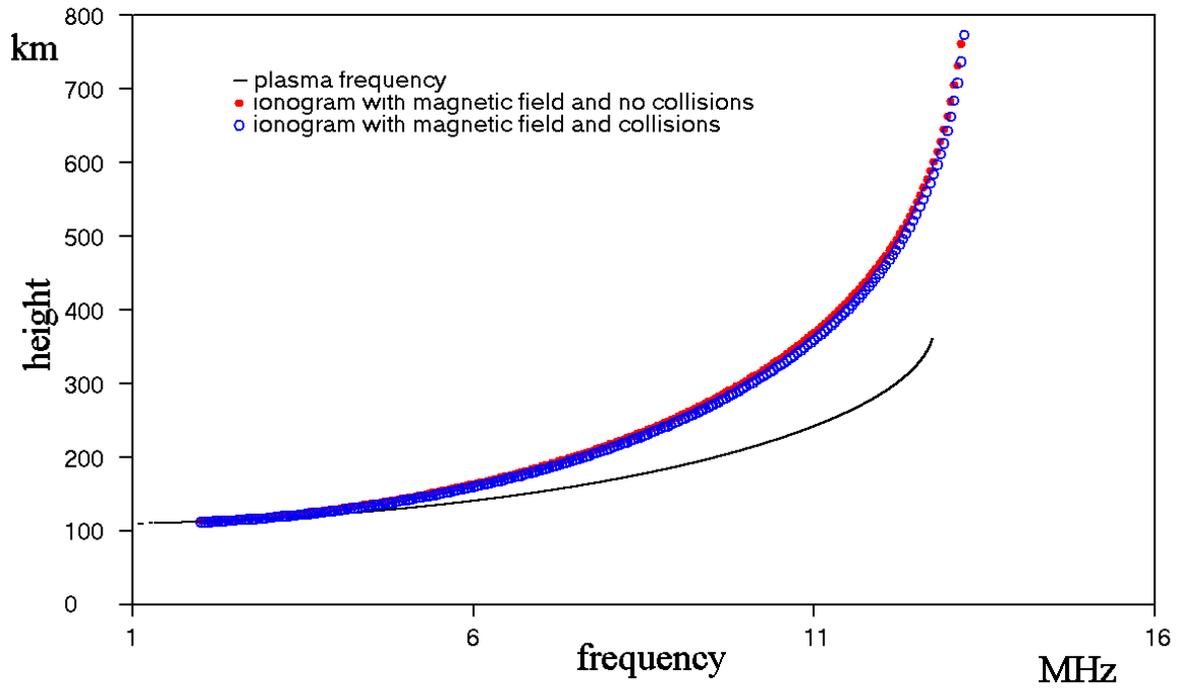

(b)

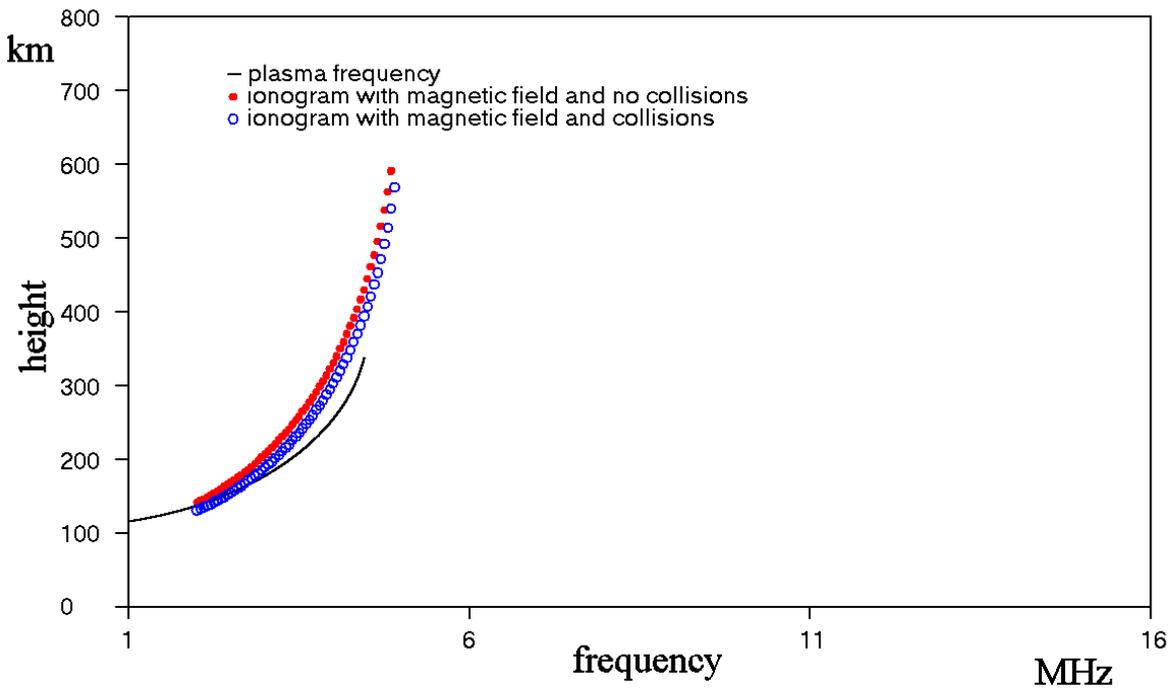

# Figure 3

**(a)**

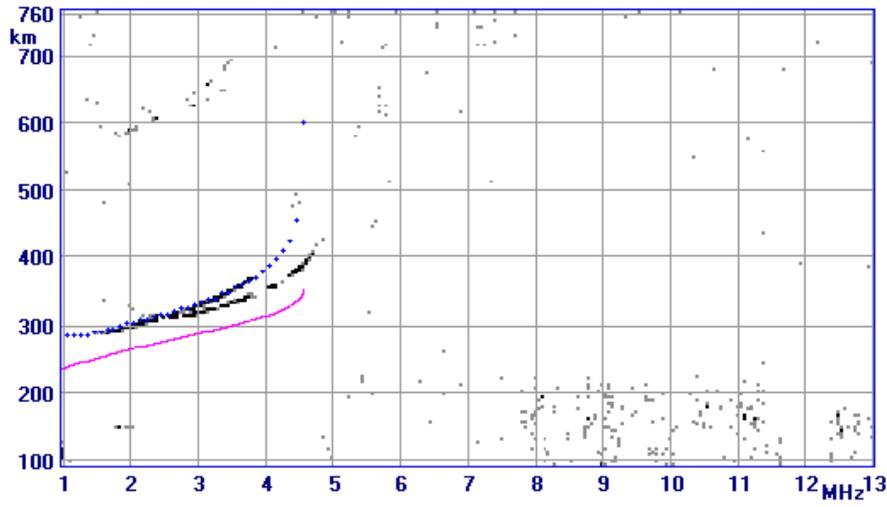

**(b)**

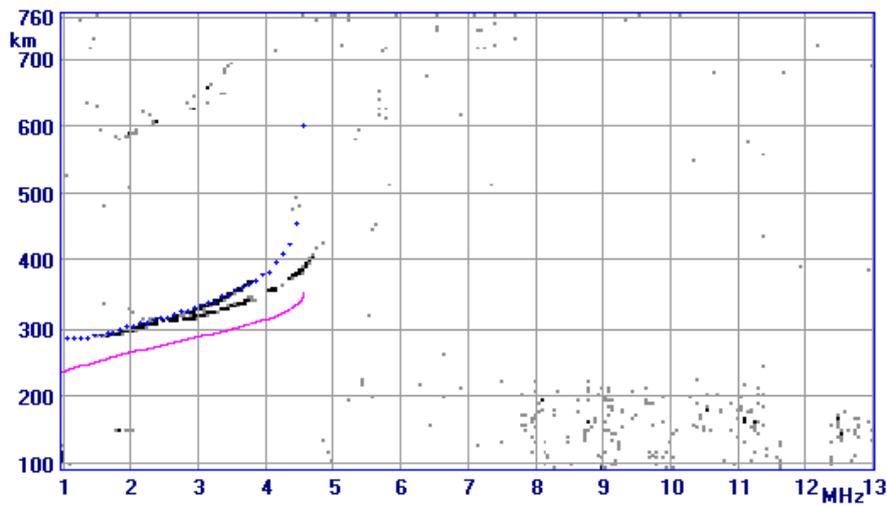

**Figure 4**

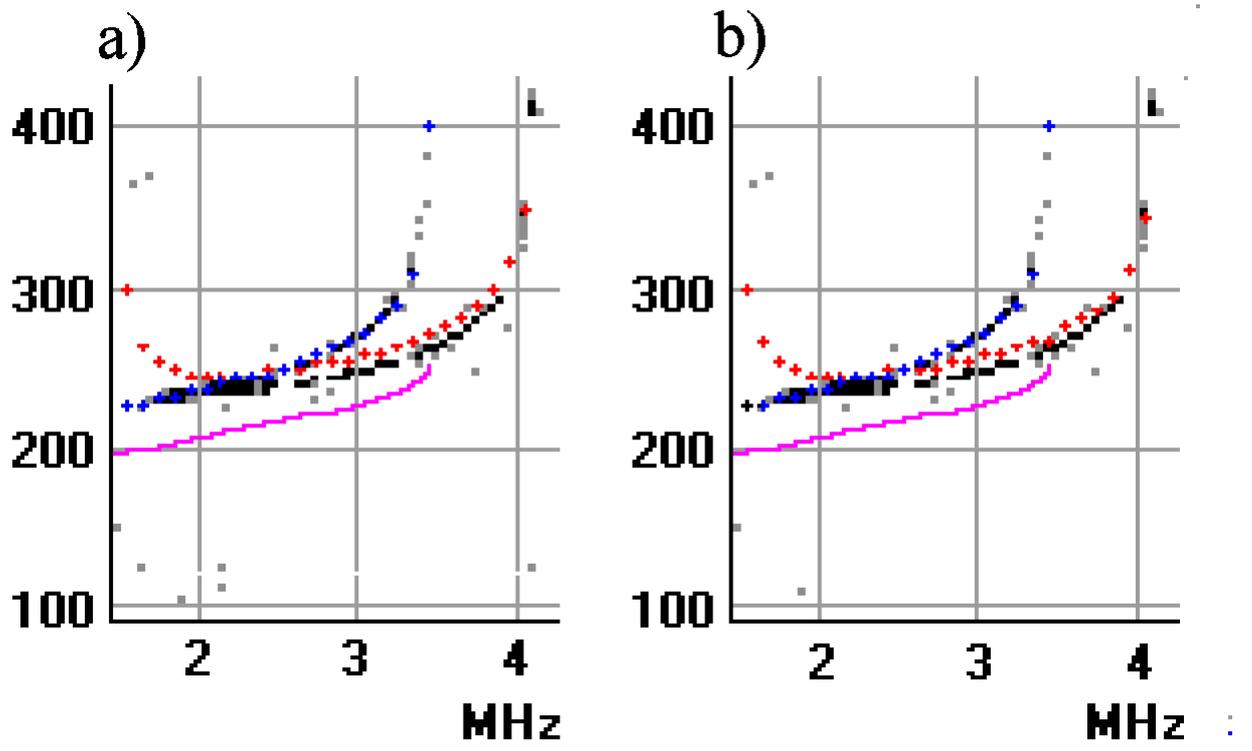